\documentclass[aps,reprint,superscriptaddress,footinbib,amsmath,amssymb,amsfonts]{revtex4-1}

\usepackage{hyperref}
\usepackage{graphicx}
\usepackage{dcolumn}
\usepackage{bm}
\usepackage{subfigure}
\usepackage{mathrsfs} 
\usepackage{epsfig}
\usepackage{epstopdf}
\usepackage{color}
\usepackage{natbib}

\newcommand{\ket}[1]{|#1\rangle}

\newcommand{\bra}[1]{\langle#1|}

\newcommand{\abs}[1]{\left|{#1}\right|}

\def\beq{\begin{eqnarray}}
\def\eeq{\end{eqnarray}}

\newcommand\redout{\bgroup\markoverwith{\textcolor{red}{\rule[.5ex]{2pt}{1.2pt}}}\ULon}
\newcommand\blueout{\bgroup\markoverwith{\textcolor{blue}{\rule[.5ex]{2pt}{1.2pt}}}\ULon}

\begin{document}

\title{Recovering quantum entanglement after its certification}

\author{Hyeon-Jin Kim}
\affiliation{Department of Physics, Korea Advanced Institute of Science and Technology, Daejeon 34141, Korea}

\author{Ji-Hyeok Jung}
\affiliation{Department of Physics, Korea Advanced Institute of Science and Technology, Daejeon 34141, Korea}

\author{Kyung-Jun Lee}
\affiliation{Department of Physics, Korea Advanced Institute of Science and Technology, Daejeon 34141, Korea}

\author{Young-Sik Ra}
\email{youngsikra@gmail.com}
\affiliation{Department of Physics, Korea Advanced Institute of Science and Technology, Daejeon 34141, Korea}

\date{\today}


\begin{abstract}
Entanglement is a crucial quantum resource with broad applications in quantum information science. For harnessing entanglement in practice, it is a prerequisite to certify the entanglement of a given quantum state. However, the certification process itself destroys the entanglement, thereby precluding further exploitation of the entanglement. Resolving this conflict, here we present a protocol that certifies the entanglement of a quantum state without complete destruction, and then, probabilistically recovers the original entanglement to provide useful entanglement for further quantum applications. We experimentally demonstrate this protocol in a photonic quantum system, and highlight its usefulness for selecting high-quality entanglement from a realistic entanglement source. Moreover, our study reveals various tradeoff relations among the physical  quantities involved in the protocol. Our results show how entanglement certification can be made compatible with subsequent quantum applications, and more importantly, be beneficial to sort entanglement for better performance in quantum technologies.
\end{abstract}

\maketitle


Entanglement---a unique feature of quantum physics---is at the heart of quantum technologies such as quantum teleportation~\cite{Ren:2017bu,Darras:2023us}, distributed quantum sensing~\cite{Guo2020,Liu:2020kr}, and quantum computing~\cite{Larsen2021, Madsen:2022jm}. To ensure its correct functioning in quantum technologies, the entanglement of a given quantum state should be certified in advance~\cite{Friis2019, Eisert:2020ko}. This entanglement certification can be classified into three different categories depending on the trust in the measurement devices of Alice and Bob. First, if both devices are trusted, one can certify the entanglement by performing quantum state tomography~\cite{James:2001ut} or an entanglement witness test~\cite{Guhne:2002em}. Second, when trusting only one device, a quantum steering test can be used~\cite{Saunders2010}, and finally, for no trust in both devices, a Bell nonlocality test can be used to certify entanglement~\cite{PhysRevLett.115.250402, PhysRevLett.115.250401, Hensen2015}.

These conventional certification protocols, however, have limitations on further exploiting the entanglement because the original entanglement is completely destroyed by quantum measurements~\cite{James:2001ut,Guhne:2002em,Saunders2010,PhysRevLett.115.250402, PhysRevLett.115.250401, Hensen2015}. In order to obtain information about a quantum state, measurements necessarily disturb the state as the measurement back action~\cite{Banaszek:2001dk}; it thus makes the resulting state no longer usable in further quantum applications. Hence, the conventional certification protocols must assume that a quantum state under a certification test, which is in turn destroyed, is identical to an unmeasured quantum state used for quantum applications~\cite{Friis2019, Eisert:2020ko}. Resolving these limitations, can a certification process be made compatible with subsequent quantum applications requiring entanglement?

Here we find an affirmative answer to this question by introducing non-projective quantum measurements---so-called weak measurements---for entanglement certification. Weak measurements can extract partial information of a quantum state without its complete destruction~\cite{Banaszek:2001dk, Lim:2014ka,Hong:2022df,White2016, PhysRevApplied.13.044008,PhysRevLett.114.250401, Hu2018,Choi:20, PhysRevA.105.032211}. Our protocol based on weak measurements provides a solution that the entanglement of a given state can be successfully certified while preserving useful entanglement for subsequent quantum applications. Interestingly, the minimum strengths of the weak measurements for successful certification depend on the level of user's trust in measurement devices, which in turn limits the amount of remaining entanglement after the certification. We find tradeoff relations of the associated quantities such as the measurement strength, the remaining entanglement, and the certification level. While the entanglement decreases by the certification, we fully recover the original entanglement by the application of reversal measurements~\cite{Kim:09, Kim2012}, which probabilistically prepares the full entanglement for subsequent quantum applications. We experimentally demonstrate this protocol in a photonic quantum system and highlight its usefulness for selecting high-quality entanglement from a realistic entanglement source.

\begin{figure}
\centerline{\includegraphics[width=0.47\textwidth]{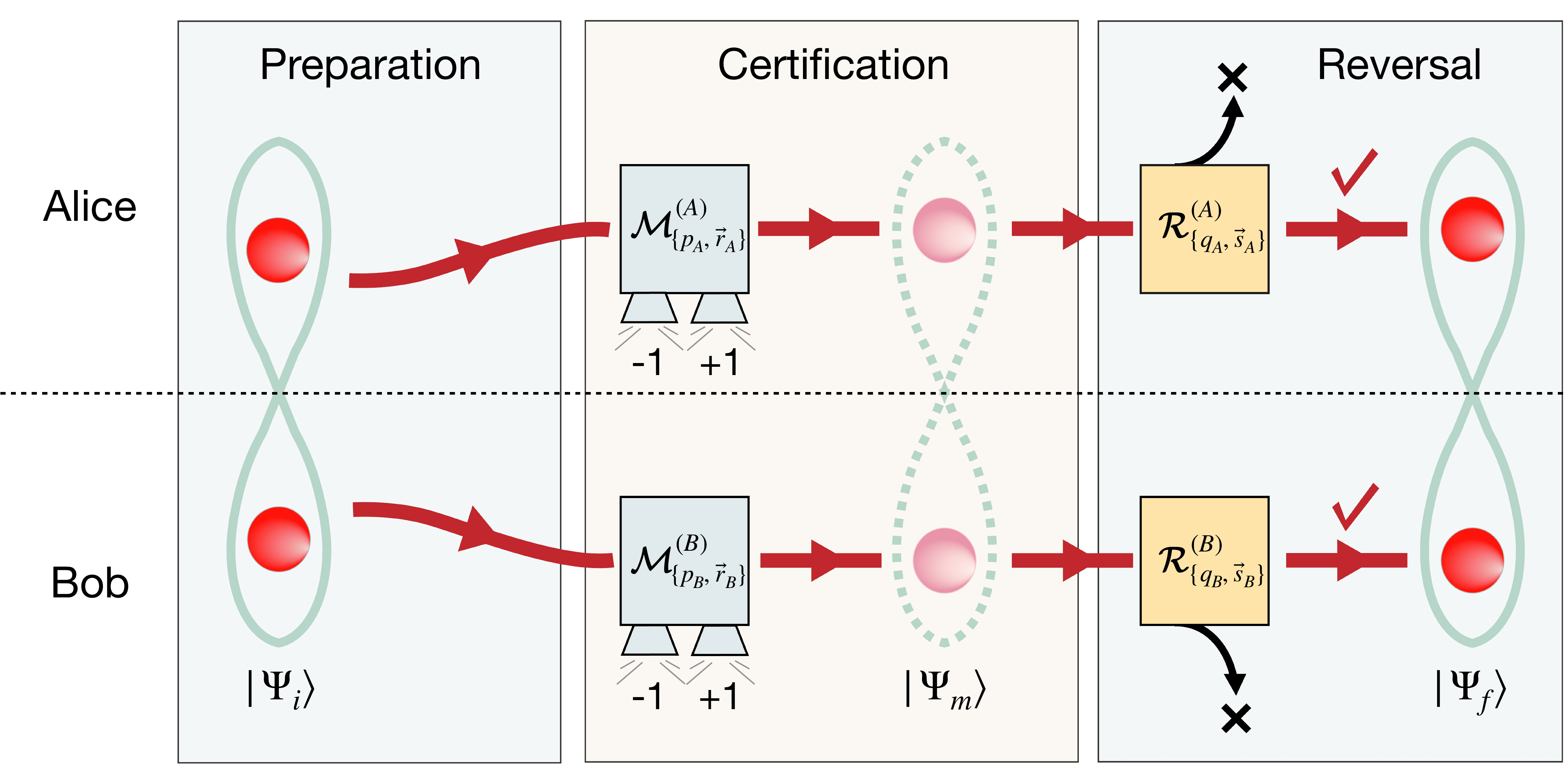}}
\caption{\label{fig1:scheme}\textbf{Conceptual scheme.} Alice ($A$) and Bob ($B$) initially share an unknown input state $\ket{\Psi_i}$. To certify entanglement of the state, they locally perform a set of weak measurements $\mathcal{M}^{(A)}_{\{p_A, \Vec{r}_A \}}$ and $\mathcal{M}^{(B)}_{\{p_B, \Vec{r}_B\}}$, where the measurement strength $p$ and direction $\Vec{r}$ can be adjusted. In this process, they obtain measurement outcomes ($\pm1$ for each), which are used for entanglement certification tests~(\ref{eq:witness},\ref{eq:steering},\ref{eq:CHSH}); as the measurement back action, the input state changes to another pure state $\ket{\Psi_m}$, which still contains some entanglement. After passing the certification tests, the original entanglement is fully recovered by reversal measurements $\mathcal{R}^{(A)}_{\{q_A, \Vec{s}_A \}}$ and $\mathcal{R}^{(B)}_{\{q_B, \Vec{s}_B \}}$, which succeeds probabilistically. As a result of these consecutive measurements, the final state becomes the original entangled state, $\ket{\Psi_f}=\ket{\Psi_i}$, and thus, it can be further harnessed for quantum technologies.
}
\end{figure}

The conceptual scheme of our protocol is described in Figure~\ref{fig1:scheme}.  To certify the entanglement of an input state $\ket{\Psi_i}$ without complete destruction, we generalize the conventional two-qubit entanglement certification protocols by introducing a non-unity strength in quantum measurement. Instead of performing projective measurements, Alice ($A$) and Bob ($B$) each performs weak measurement $\mathcal{M}^{(k)}_{\{p_k, \Vec{r}_k \}}$ ($k=A, B$), where $p_k$ and $\Vec{r}_k$ represent the measurement strength and direction, respectively. The measurement strength $p_k$ ranges from 0 (no measurement) to 1 (projective measurement). This weak measurement consists of measurement operators $\{ \hat{M}^{(k)}_{l_k |\{p_k,\Vec{r}_k\}}   \}$ with the two possible outcomes $l_k\in\{+1, -1\}$, where $\hat{M}^{(k)}_{\pm|\{p_k,\Vec{r}_k\}} = \sqrt{(1+ p_k)/2}~\hat{\Pi}^{(k)}_{\pm|\Vec{r}_k} + \sqrt{(1- p_k)/2}~\hat{\Pi}^{(k)}_{\mp|\Vec{r}_k}$, and $\hat{\Pi}^{(k)}_{\pm|\Vec{r}_k} = \frac{1}{2}(\hat{I}^{(k)} \pm \Vec{r}_k\cdot\Vec{\hat{\sigma}}^{(k)})$ is the projection operator to the direction $\Vec{r}_k$ ($\hat{I}$: an identity operator, $\Vec{\sigma}$: Pauli operators $(\hat{\sigma}_x, \hat{\sigma}_y, \hat{\sigma}_z)$). We can then define a generalized observable $\hat{\mu}^{(k)}_{\{p_k, \Vec{r}_k\}}$ associated with the weak measurement: $\hat{\mu}^{(k)}_{\{p_k, \Vec{r}_k\}} = \sum_{l_k=\pm1} l_k ~ \hat{M}_{l_k |\{p_k,\Vec{r}_k\}}^{(k)}{}^{\dag}~\hat{M}_{l_k |\{p_k,\Vec{r}_k\}}^{(k)} $.
Note that $\langle \hat{\mu}^{(k)}_{\{p_k, \Vec{r}_k\}} \rangle$ gives the expectation value of the measurement outcomes, and $\hat{\mu}^{(k)}_{\{p_k, \Vec{r}_k\}}$ is related to the Pauli observable $\hat{\sigma}^{(k)}_{\Vec{r}_k}=\Vec{r}_k \cdot \Vec{\hat{\sigma}}^{(k)}$ (which is used for projective measurements) via
\begin{equation}
\hat{\mu}^{(k)}_{\{p_k, \Vec{r}_k\}} = p_k \hat{\sigma}^{(k)}_{\Vec{r}_k}.
\label{eq:trusted}
\end{equation}
This relation allows us to adapt conventional certification tests for weak measurements in a simple way. When a measurement device is trusted (meaning that the device is fully characterized), $\hat{\sigma}^{(k)}_{\Vec{r}_k}$ in certification tests can be replaced by $\frac{1}{p_k} \hat{\mu}^{(k)}_{\{p_k, \Vec{r}_k\}}$ using Eq.(\ref{eq:trusted}), but without such trust, $\hat{\sigma}^{(k)}_{\Vec{r}_k}$ should be directly replaced by $\hat{\mu}^{(k)}_{\{p_k, \Vec{r}_k\}}$.

We consider three conventional certification tests assuming different levels of trust in measurement devices: witness, quantum steering, and Bell nonlocality. Without loss of generality, we choose a target state of $\ket{\Psi_i} = \frac{1}{\sqrt{2}}(\ket{++}+\ket{--})$ for the certification. At first, when Alice and Bob trust their measurement devices, we use the witness test~\cite{Guhne:2002em}:
\begin{equation}
W = \frac{1}{4} - \frac{1}{4} \sum_{\Vec{r}\in\{\Vec{x},\Vec{y},\Vec{z}\}} w_{\Vec{r}}~ \langle \hat{\sigma}^{(A)}_{\Vec{r}}\hat{\sigma}^{(B)}_{\Vec{r}}\rangle  < 0
\label{eq:witness0}
\end{equation}
with the weights of $w_{\Vec{x}} = - w_{\Vec{y}} = w_{\Vec{z}} = 1$. Since the devices at $A$ and $B$ are fully characterized, we can make use of Eq.(\ref{eq:trusted}) to express the witness by the generalized observables:
\begin{equation}
W = \frac{1}{4} - {1 \over {4p_A p_B}}\sum_{\Vec{r}\in\{\Vec{x},\Vec{y},\Vec{z}\}} w_{\Vec{r}}~ \langle \hat{\mu}^{(A)}_{\{p_A, \Vec{r}\}} \hat{\mu}^{(B)}_{\{p_B, \Vec{r}\}}  \rangle < 0,
\label{eq:witness}
\end{equation}
where the joint expectation value $\langle \hat{\mu}^{(A)}_{\{p_A, \Vec{r}\}} \hat{\mu}^{(B)}_{\{p_B, \Vec{r}\}}  \rangle$ is obtained by the measurement outcomes $l_A$ and $l_B$. In the case of the witness test, weak measurements of any non-zero measurement strengths can be used to certify entanglement.
Next, when the device at Bob is only trusted, a quantum steering test is employed for entanglement certification. One can certify entanglement by showing that Alice can steer Bob's quantum state, which is represented by~\cite{Saunders2010}
\begin{equation}
S_3 = \frac{1}{3p_B}\sum_{\Vec{r}\in\{\Vec{x},\Vec{y},\Vec{z}\}} w_{\Vec{r}}~ \langle \hat{\mu}^{(A)}_{\{p_A, \Vec{r}\}} \hat{\mu}^{(B)}_{\{p_B, \Vec{r}\}} \rangle > \frac{1}{\sqrt{3}}.
\label{eq:steering}
\end{equation}
For obtaining the result, Eq. (\ref{eq:trusted}) has been applied only for the trusted device of Bob. Then, the minimum requirement on the measurement strength is $p_A>1/\sqrt{3}$ for Alice and $p_B>0$ for Bob. Similarly, quantum steering in the reverse direction, where Bob steers Alice's state, can be constructed by exchanging $A$ and $B$. 
Finally, for untrusted devices on both sides, a Bell nonlocality test is used, which is the most stringent test for entanglement certification~\cite{RevModPhys.86.419}:
\begin{equation}
S =\abs{\langle \hat{\alpha}_{1}\hat{\beta}_{1}\rangle + \langle \hat{\alpha}_{1}\hat{\beta}_{2}\rangle + \langle \hat{\alpha}_{2}\hat{\beta}_{1}\rangle - \langle \hat{\alpha}_{2}\hat{\beta}_{2}\rangle} > 2.
\label{eq:CHSH}
\end{equation}
Here we cannot use Eq.~(\ref{eq:trusted}) because both devices are untrusted, and thus, the observables are directly replaced as $\hat{\alpha}_{1} = \hat{\mu}_{\{p_A, \Vec{z}\}}^{(A)}$, $\hat{\alpha}_{2} = \hat{\mu}_{\{p_A, \Vec{x}\}}^{(A)}$, $\hat{\beta}_{1} = \hat{\mu}_{\{p_B,{1 \over {\sqrt{2}}} ({\Vec{z}+\Vec{x})}\}}^{(B)}$, and $\hat{\beta}_{2} = \hat{\mu}_{\{p_B, {1 \over \sqrt{2}} ({\Vec{z}-\Vec{x}})\}}^{(B)}$. The required condition for the measurement strengths is $p_A p_B > 1/\sqrt{2}$, which is still achievable by weak measurements.

\begin{figure*}
\centerline{\includegraphics[width=0.8\textwidth]{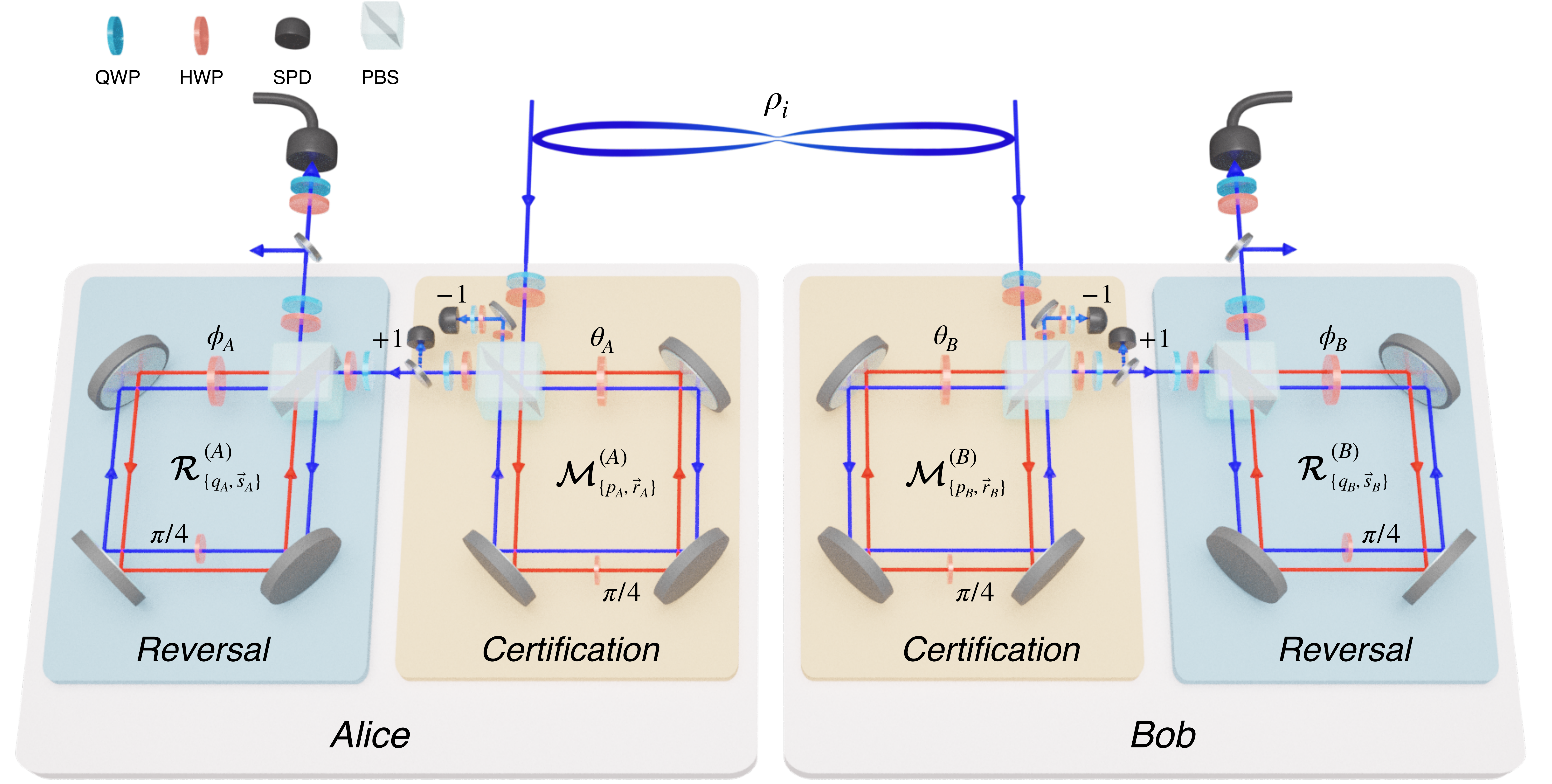}}
\caption{\label{fig2:setup}\textbf{Experimental setup.} A two-photon entangled state $\rho_i$ is distributed to Alice (A) and Bob (B). For certifying entanglement of the state, they locally  perform weak measurement $\mathcal{M}^{(k)}_{\{p_k, \Vec{r}_k \}}$ ($k=A,B$), which is implemented by a Sagnac interferometer. For each side, the measurement strength is adjusted by rotating the half wave plate (HWP) angle $\theta_{k}$, and the measurement basis is controlled by the two pairs of a HWP and a quarter wave plate (QWP) before and after a Sagnac interferometer. Measurement outcome  ($-1$, $+1$) is obtained by a click on a single-photon detector (SPD), which is used for entanglement certification tests given in (\ref{eq:witness},\ref{eq:steering},\ref{eq:CHSH}). The reversal measurement $\mathcal{R}^{(k)}_{\{q_k, \Vec{s}_k \}}$ is similarly implemented by another Sagnac interferometer, where the measurement strength is set to be the same as the weak measurement (i.e., $\phi_k = \theta_k$) for recovering entanglement. More details about the weak and reversal measurements are described in Methods. The recovery of the initial entanglement is verified by performing quantum state tomography on the final state. PBS stands for a polarizing beam splitter.
}
\end{figure*}

The next step is to recover the original entanglement from the partially disturbed state as a result of the entanglement certification. For this purpose, we employ reversal measurements. More specifically, from the disturbed state $\ket{\Psi_m} = {{1}\over{\sqrt{\mathcal{N}}}}  ( \hat{M}^{(A)}_{l_A | \{p_A,\Vec{r}_A\}} \otimes \hat{M}^{(B)}_{l_B |\{p_B,\Vec{r}_B\}}) \ket{\Psi_i}$ ($\mathcal{N}$: the normalization constant, $l_{A}$, $l_{B}$: measurement outcomes), Alice and Bob perform reversal measurements $\mathcal{R}^{(A)}_{\{q_A, \Vec{s}_A \}}$ and $\mathcal{R}^{(B)}_{\{q_B, \Vec{s}_B\}}$, where the associated measurement operators are $\hat{R}^{(k)}_{\pm|\{q_k,\Vec{s}_k\}} = \sqrt{(1- q_k)/2}~\hat{\Pi}^{(k)}_{\pm|\Vec{s}_k} + \sqrt{(1+ q_k)/2}~\hat{\Pi}^{(k)}_{\mp|\Vec{s}_k}$ for the measurement outcome $\pm 1$. For successful recovery, the reversal measurements should satisfy the following conditions: $q_A=p_A, \Vec{s}_A=\Vec{r}_A, q_B=p_B, \Vec{s}_B=\Vec{r}_B$. When the outcomes of the initial  and the reversal measurements are identical for each of Alice and Bob, the initial quantum state is finally recovered to $\ket{\Psi_f}=\frac{1}{4} \sqrt{(1-p_A^2)(1-p_B^2)}\ket{\Psi_i}$. Here the normalization factor is interpreted as the probability of successful recovery, and if we take into accounts the four possible cases of the identical outcomes, the total probability of recovery---called \textit{reversibility}---becomes $R=\frac{1}{4}(1-p_A^2)(1-p_B^2)$. The reversibility depends only on the measurement strengths of Alice and Bob.

We experimentally demonstrate this protocol in a photonic quantum system. Figure~\ref{fig2:setup} describes our experimental setup of the entanglement certification and recovery. We initially generate a two-photon entangled state, $\ket{\Psi_i} = \frac{1}{\sqrt{2}}(\ket{HH}+\ket{VV})$, by using spontaneous parametric down-conversion (SPDC) in a nonlinear crystal (see Methods for details). To certify the entanglement of the generated state without full destruction, we employ weak measurements based on Sagnac interferometers. The measurement strength is controlled by rotating the HWP angle $\theta_k$, giving the measurement strength of $p_k = \abs{\cos{4\theta_k}}$, and the basis is controlled by the two pairs of HWP and QWP. For various measurement strengths, we have characterized the weak measurements by performing quantum process tomography, which agrees well with the ideal operations (see Supplementary Information).

\begin{figure*}
\centerline{\includegraphics[width=0.9\textwidth]{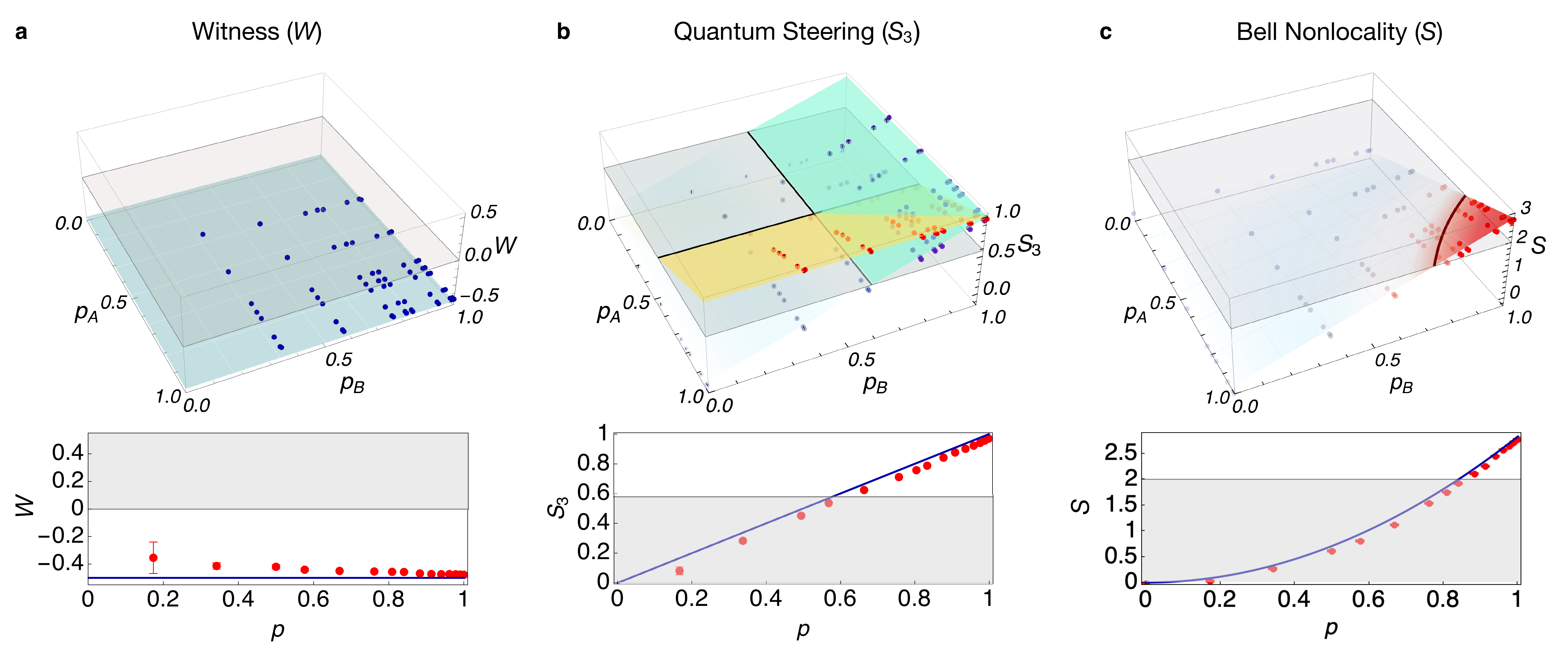}}
\caption{\label{fig3:certification}\textbf{Entanglement certification by weak measurements.} Entanglement of the initial state is certified by (a) witness test, (b) quantum steering $S_3$, and (c) Bell nonlocality $S$ with adjustable measurement strengths $p_A$ and $p_B$. The grey horizontal planes are the bounds for entanglement certification (entangled if $W < 0$, $S_3 > 1/\sqrt{3} $, or $S>2$). The other surfaces show the theoretical plots by the ideal conditions. In (b), there are two theoretical plots: the yellow is for the quantum steering from Alice to Bob, and the cyan is for the quantum steering from Bob to Alice. Markers are experimental data, which are placed close to the theoretical plots. The insets below the 3D figures are projections along $p=p_A=p_B$, where the white area represents the successful entanglement certification. The error bars are the standard deviations, obtained by more than 17 repeated experiments.
}
\end{figure*}

Using the weak measurements, we conduct the three different entanglement certification tests in (\ref{eq:witness},\ref{eq:steering},\ref{eq:CHSH}). First, we certify the entanglement of the experimentally generated state using the witness test (\ref{eq:witness}), whose result is shown in Fig.~\ref{fig3:certification}(a). The entanglement is certified in the full range of non-zero measurement strengths of $p_A$ and $p_B$, exhibiting $W<0$; the full knowledge about the weak measurements has made it possible to compensate the non-unity measurement strengths in the certification test. Next, we certify the entanglement using quantum steering, as presented in Fig.~\ref{fig3:certification}(b). Entanglement can be certified by Alice's steering on the Bob's state if $p_A > 1/\sqrt{3}$ or similarly done in the opposite direction if $p_B > 1/\sqrt{3}$. Interestingly, one finds the asymmetric features of quantum steering with respect to measurement strengths while such features have usually been attributed to a quantum state~\cite{Handchen2012, PhysRevLett.112.200402}. Within the certification range of the measurement strengths, quantum steering is achieved in both directions if $p_A > 1/\sqrt{3}$ and $p_B > 1/\sqrt{3}$, but, for the other case, quantum steering can be done only in one direction. Finally, we employ a Bell nonlocality test for entanglement certification, as shown in Fig.~\ref{fig3:certification}(c). Without assuming any trusts in the measurement devices, the entanglement can be certified in the range of measurement strengths satisfying $p_A p_B > 1/\sqrt{2}$.

\begin{figure*}
\centerline{\includegraphics[width=0.9\textwidth]{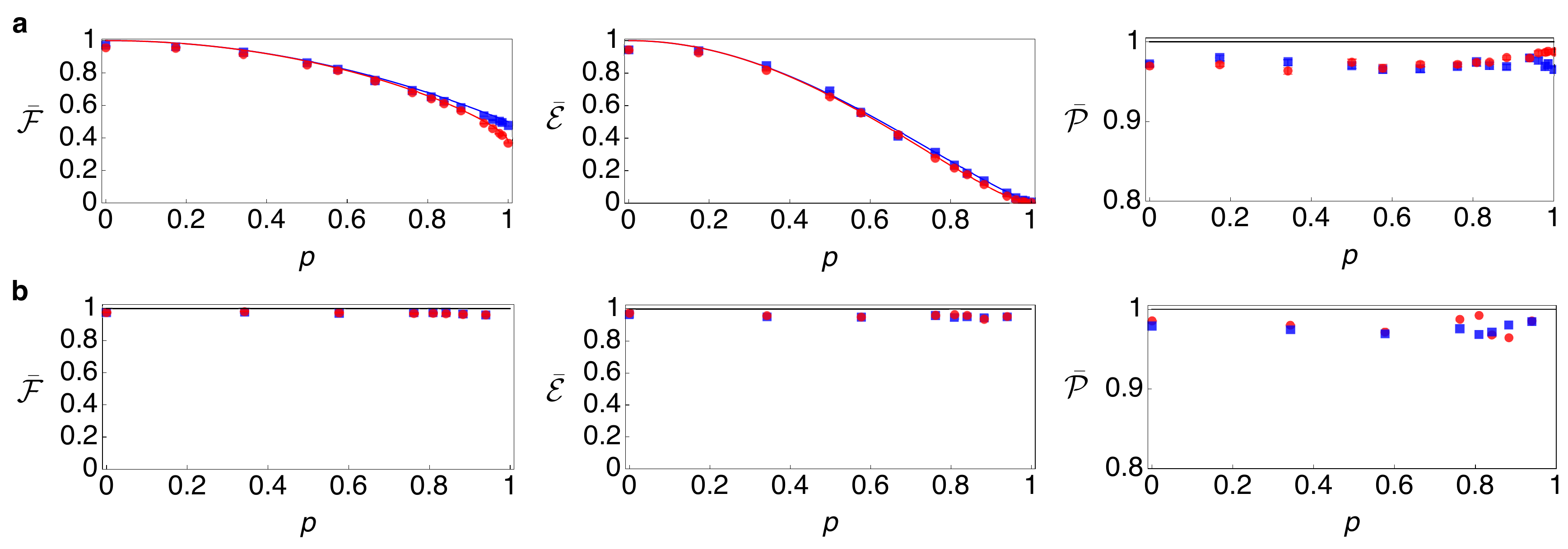}}
\caption{\label{fig4:recover} \textbf{Recovery of quantum entanglement.} Properties of the quantum states (a) before and (b) after applying reversal measurements (average fidelity $\bar{\cal{F}}$, average entanglement of formation $\bar{\cal{E}}$, and average purity $\bar{\cal{P}}$). The measurement strengths for Alice and Bob are set to be identical ($p=p_A=p_B$). Dots and lines are the experimental data and the ideal theory graphs, respectively, where blue is for witness and quantum steering, and red is for Bell nonlocality. Black lines are used if their theory graphs are identical. Error bars denote one standard deviation.
}
\end{figure*}

As a result of the certification, the initial quantum state is partially disturbed. To investigate the characteristics of the disturbed state, we consider an average quantity $\bar{\cal{Q}}$ over all possible measurement outcomes $l_A$ and $l_B$ by different sets of measurement directions $({\Vec{r}_A}, {\Vec{r}_B})$ for entanglement certification:
\begin{equation}
\bar{\cal{Q}} = \frac{1}{D} \sum_{(\Vec{r}_A,\Vec{r}_B)}^D \sum_{l_A=\pm1}\sum_{l_B=\pm1} P(l_A,l_B | \Vec{r}_A,\Vec{r}_B)~ {\cal{Q}} \left[\rho_m \right],  
\label{eq:average}
\end{equation}
where $P(l_A,l_B | \Vec{r}_A,\Vec{r}_B)$ is the probability to obtain the measurement outcomes $l_A$ and $l_B$ for the measurement directions of $\Vec{r}_A$ and $\Vec{r}_B$, $\rho_m$ is the disturbed quantum state associated with the outcomes, and $\cal{Q}$ is the quantity of interest. For the witness and the steering tests, the required measurement directions are $(\Vec{x},\Vec{x})$, $(\Vec{y},\Vec{y})$, $(\Vec{z},\Vec{z})$, thus $D=3$, and for the Bell nonlocality test, the directions are $(\Vec{z},\frac{\Vec{z}+\Vec{x}}{\sqrt{2}})$, $(\Vec{z},\frac{\Vec{z}-\Vec{x}}{\sqrt{2}})$, $(\Vec{x},\frac{\Vec{z}+\Vec{x}}{\sqrt{2}})$, $(\Vec{x},\frac{\Vec{z}-\Vec{x}}{\sqrt{2}})$, thus $D=4$. We then study the following average quantities: the average fidelity $\bar{\cal{F}}$ (the fidelity with the initial state), the average entanglement of formation $\bar{\cal{E}}$, and the average purity $\bar{\cal{P}}$. The results are plotted in Fig.~\ref{fig4:recover}(a), where the full data before averaging over different measurement directions are provided in Supplementary Information. As expected, the fidelity and the entanglement decrease as the measurement strength increases. However, the purity is unaffected because the weak measurements do not introduce noise. The observed reduction of entanglement is attributed to the imbalance of probability amplitudes in the quantum state rather than generation of a mixed state, suggesting that appropriate quantum operations can recover the original entanglement.

\begin{figure}
\centerline{\includegraphics[width=0.35\textwidth]{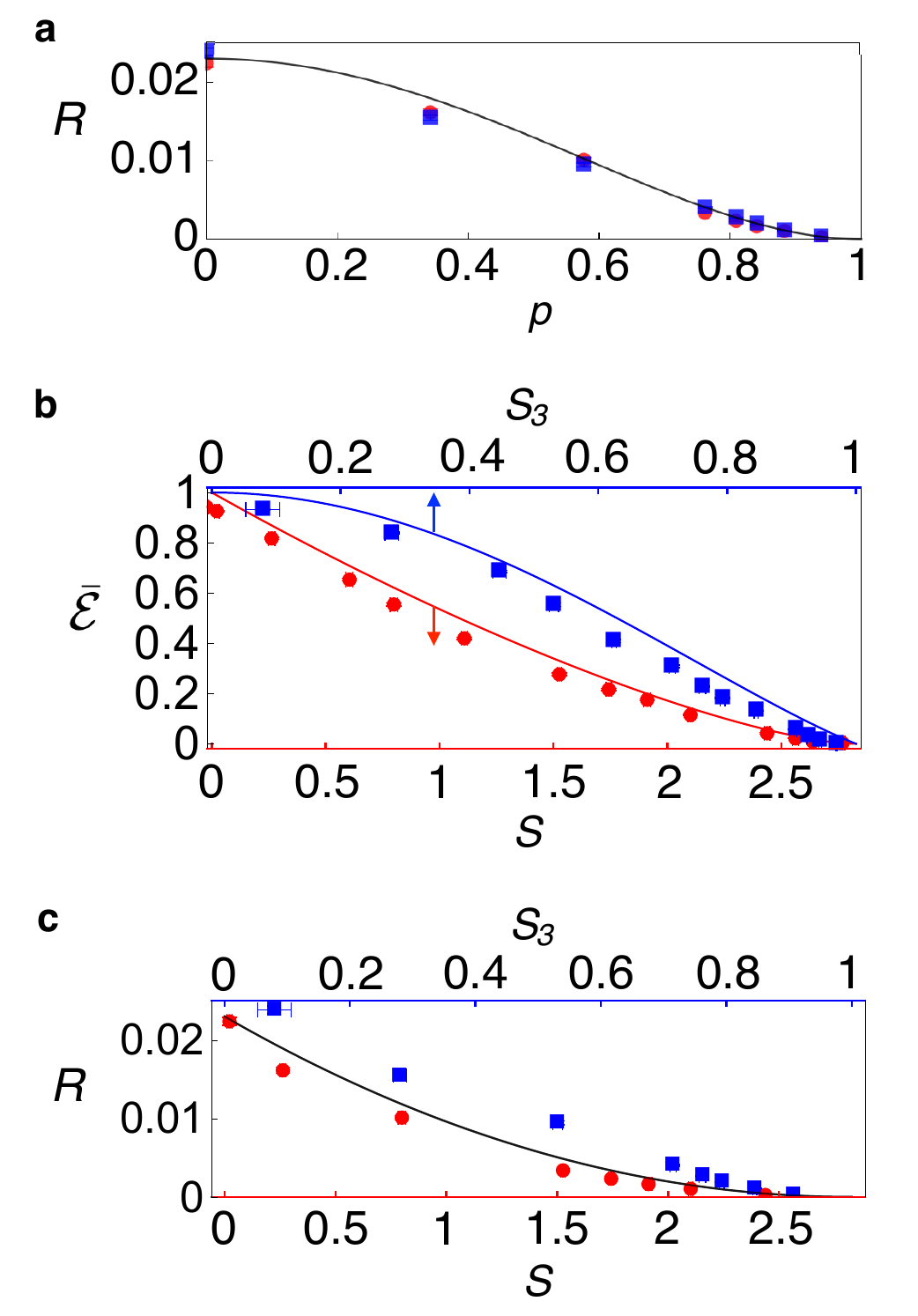}}
\caption{\label{fig5:tradeoff}\textbf{Tradeoff relations.} Relations of various quantities involved in the protocol (measurement strength $p$, reversibility $R$, Bell nonlocality $S$, quantum steering $S_3$, and average entanglement of formation $\bar{\cal{E}}$).
Dots and lines are the experimental data and the ideal theory graphs, respectively, where red is for Bell nonlocality, and blue is for quantum steering. 
The theory graph of reversibility (black line) assumes the average optical loss of 79 $\%$ at each of Alice and Bob. Error bars denote one standard deviation.
}
\end{figure}

To fully recover the original entanglement, we perform reversal measurements on the disturbed state. Figure~\ref{fig4:recover}(b) shows the result of recovery: the final state exhibits near-unity values of fidelity, entanglement, and purity. For the demonstration purpose, we have applied the reversal measurement only for $+1$ outcome of the weak measurement, but one can similarly apply reversal measurement for the other outcome $-1$ which will just increase the success probability. The recovery process is probabilistic, where the reversibility (i.e., the success probability) $R$ decreases as the measurement strength increases. This tradeoff relation is plotted in Fig.~\ref{fig5:tradeoff}(a), together with other tradeoff relations of relevant quantities as shown in Fig.~\ref{fig5:tradeoff}(b-c).

\begin{figure}
\centerline{\includegraphics[width=0.37\textwidth]{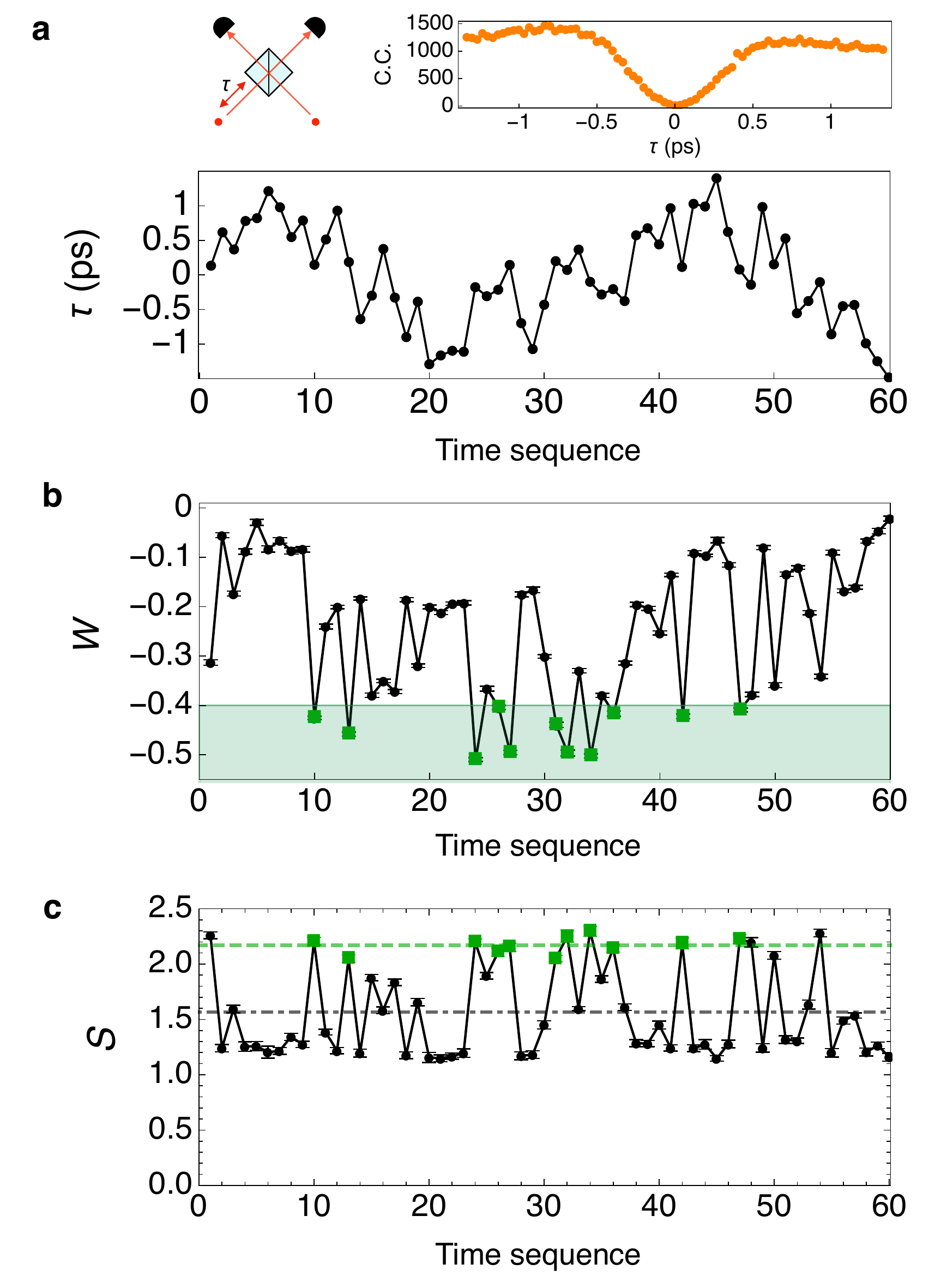}}
\caption{\label{fig6:selection}\textbf{Selection of high-quality entanglement from an entanglement source.} (a) A random drift of a photon's arrival time $\tau$ in a Hong-Ou-Mandel type interferometer~\cite{Kim2003}. The inset shows the interferometer with $\tau$ and the measured interference data. For simulating the drift, we use random numbers generated from the Ornstein-Uhlenbeck process. (b) Witness values obtained in the certification process. For selecting high-quality entanglement, we choose the cases of $W < -0.4$, which are represented by green squares. (c) Bell nonlocality test as a subsequent quantum application. With the information from the certification process, the Bell nonlocality test succeeds, resulting in the average value of $S = 2.17 \pm 0.01 > 2$ (green dashed line); on the other hand, without the prior certification, the test fails by giving $S = 1.57 \pm 0.01 < 2$ (grey dot-dashed line). Error bars of $W$ and $S$ denote one standard deviation.
}
\end{figure}

Finally, we illustrate the usefulness of our protocol in real applications of an entanglement source to quantum technologies. In practice---because a perfectly entangled state is not always generated---, the quality of an entanglement source should be monitored continuously~\cite{Friis2019,Eisert:2020ko}. Our protocol can monitor the quality of an entanglement source and select sufficiently high-quality entanglement for subsequent quantum applications. To simulate an entanglement source with time-varying decoherence, we consider a random drift of the arrival time of a photon at a beam splitter for generating entanglement~\cite{Kim2003}, as depicted in Fig.~\ref{fig6:selection}(a). The resulting state is described by a mixed state $\rho_{\rm mix} (t) = (1-\gamma(t)) \ket{\Psi_i}\bra{\Psi_i}+\frac{1}{2} \gamma(t) (\ket{HH}\bra{HH}+\ket{VV}\bra{VV})$, where $\gamma(t)$ accounts for the decoherence degree. In the certification process, we monitor the time-varying quality of entanglement by observing the witness $W$, which can provide a lower bound on the amount of entanglement~\cite{Eisert:2007us,Guhne:2007vo}. $W$ obtained in the certification process is plotted in Fig.~\ref{fig6:selection}(b), and we will select the cases of $W < - 0.4$ for high-quality entanglement. Finally, the reversal measurement is applied, which recovers the original entanglement for further quantum applications. As an example of such quantum applications, we adopt a Bell nonlocality test because nonlocality usually serves as the fundamental quantum resource behind various quantum technologies~\cite{Reichardt:2013vg,Bierhorst:2018fs,Liu:2022fg,Zhang2022}. The experimental results are shown in Fig.~\ref{fig6:selection}(c). Without the information of $W$, the Bell test would have failed by giving $S = 1.57 \pm 0.01 < 2$, but with the information of $W$ (by selecting $W < - 0.4$), the Bell test succeeds by exhibiting $S = 2.17 \pm 0.01 > 2$. This example demonstrates how our protocol can be beneficial to select high-quality entanglement for a better performance in quantum technologies.

In conclusion, we propose and demonstrate a protocol that certifies the entanglement of a quantum state without fully destroying it, and then, recovers the original entanglement for subsequent quantum applications. Our work shows how entanglement certification can be made compatible with subsequent quantum applications, thereby lifting the standard assumption (identical quantum states for a certification test and a quantum application) required in conventional certification protocols~\cite{Friis2019, Eisert:2020ko}. Our protocol generalizes entanglement certification by incorporating non-destructive quantum measurements, which has been applied for various certification tests assuming different levels of trusts in the measurement devices~\cite{Guhne:2002em,Saunders2010,RevModPhys.86.419}. We have shown that our generalized protocol can successfully certify the entanglement by preserving useful entanglement, where the following reversal measurement fully recovers the original entanglement in a probabilistic way. Notably, such generalization reveals profound tradeoff relations about quantum measurement and quantum entanglement, stimulating further studies on information balances focused on entanglement~\cite{Banaszek:2001dk,Lim:2014ka,Hong:2022df}. From a practical perspective, our protocol is beneficial for enhancing the performance of quantum technologies by selecting high-quality entanglement from a realistic entanglement source. Our certification protocol may find broad applications in entanglement-based quantum technologies~\cite{Ren:2017bu,Darras:2023us,Guo2020,Liu:2020kr,Larsen2021,Madsen:2022jm,Reichardt:2013vg,Bierhorst:2018fs,Liu:2022fg,Zhang2022}, which is applicable to other quantum systems as well (e.g. superconductors~\cite{White2016} and trapped ions~\cite{Pan:2020tc}).

\section{Methods}
\textbf{Entanglement generation.} We experimentally generate an entangled photon pair via type-II SPDC by pumping a 10-mm-thick periodically-polled KTP crystal using a 405 nm diode laser. The spectrum of each photon is filtered by a bandpass filter of 3 nm full width half maximum centered at 810 nm. Each photon then enters a single-mode fiber for spatial mode filtering. The two photons, each exiting from a single mode fiber, arrive at a PBS simultaneously, resulting in the entangled state of $\ket{\Psi_i} = \frac{1}{\sqrt{2}}(\ket{HH}+\ket{VV})$~\cite{Jeong:2016eo,Lee:16}.

\textbf{Weak and reversal measurements.}
To implement weak and reversal measurements, we construct Sagnac interferometers described in Fig.~\ref{fig2:setup}. For each measurement, a pair of HWP and QWP at each of input and output of an interferometer implement a measurement basis change $\hat{U}_{\vec{r}_k}$, and the HWP at an angle of $\theta_k$ (or $\phi_k$) controls the measurement strength $p_k = \abs{\cos{4\theta_k}}$ (or $\abs{\cos{4\phi_k}}$). More specifically, in the PBS, a beam with $H$ polarization is transmitted while $V$ polarization is reflected, resulting in two beams propagating in opposite directions. Inside the interferometer, both beams go through the HWP at $\theta_k$ (or $\phi_k$), but only one beam goes through the HWP at $\pi/4$. After the two beams are overlapped together at the PBS, SPDs at the output detect a single photon, producing measurement outcomes $\pm 1$ depending on the detector click position. The corresponding measurement operators are $\hat{M}^{(k)}_{\pm | \{ p_k,\vec{r}_k \} } = \sqrt{(1\pm p_k)/2} ~ \hat{U}^{\dagger}_{\vec{r}_k} \ket{V}\bra{V} \hat{U}_{\vec{r}_k} ~+~ \sqrt{(1\mp p_k)/2}  ~ \hat{U}^{\dagger}_{\vec{r}_k} \ket{H}\bra{H}\hat{U}_{\vec{r}_k}$ for a weak measurement and $\hat{R}^{(k)}_{\pm | \{ p_k,\vec{r}_k \} } = \sqrt{(1\mp p_k)/2}  ~ \hat{U}^{\dagger}_{\vec{r}_k} \ket{V}\bra{V} \hat{U}_{\vec{r}_k} ~+~ \sqrt{(1\pm p_k)/2}  ~ \hat{U}^{\dagger}_{\vec{r}_k} \ket{H}\bra{H}\hat{U}_{\vec{r}_k}$ for a reversal measurement.

\section{Acknowledgment}
We thank Y.-C. Jeong, H.-T. Lim, and S.-W. Lee for fruitful discussions. This work was supported by the Ministry of Science and ICT (MSIT) of Korea (NRF-2020M3E4A1080028, NRF-2022R1A2C2006179) under the Information Technology Research Center (ITRC) support program (IITP-2023-2020-0-01606) and Institute of Information \& Communications Technology Planning \& Evaluation (IITP) grant (No. 2022-0-01029, Atomic ensemble based quantum memory).


\providecommand{\noopsort}[1]{}\providecommand{\singleletter}[1]{#1}%

\end{document}